\documentclass[%
reprint,
 amsmath,amssymb,nofootinbib,
aps,
]{revtex4-2}

\usepackage{float}
\usepackage{graphicx}
\usepackage{dcolumn}
\usepackage{bm}
\usepackage[normalem]{ulem}
\usepackage{hyperref}

\usepackage{physics}
\usepackage{xcolor}

\begin{document}


\title{Clarifications on black/white hole metastable relics as dark matter}

\author{Aurélien Barrau}
 \email{barrau@lpsc.in2p3.fr}
 \affiliation{
 Laboratoire de Physique Subatomique et de Cosmologie, Univ. Grenoble Alpes, CNRS/IN2P3\\ 53 avenue des Martyrs, 38026 Grenoble Cedex, France
}
\author{Ruth Durrer}
\email{ruth.durrer@unige.ch}
\affiliation{Département de Physique Théorique, Université de Genève,\\24 quai Ernest Ansermet, 1211 Genève 4, Switzerland}
\author{Margaux Gajocha}
\email{margaux.gajocha@ens-lyon.fr}
\affiliation{
 Laboratoire de Physique Subatomique et de Cosmologie, Univ. Grenoble Alpes, CNRS/IN2P3\\ 53 avenue des Martyrs, 38026 Grenoble Cedex, France
}
\author{Killian Martineau}
 \email{martineau@lpsc.in2p3.fr}
 \affiliation{
 Laboratoire de Physique Subatomique et de Cosmologie, Univ. Grenoble Alpes, CNRS/IN2P3\\ 53 avenue des Martyrs, 38026 Grenoble Cedex, France
}
 \author{Martin Teuscher}
\email{teuscher@lpsc.in2p3.fr}
\affiliation{
 Laboratoire de Physique Subatomique et de Cosmologie, Univ. Grenoble Alpes, CNRS/IN2P3\\ 53 avenue des Martyrs, 38026 Grenoble Cedex, France
}
\author{Jamy-Jayme Thézier}
\email{jamyjayme.thezier@gmail.com}
\affiliation{
 Laboratoire de Physique Subatomique et de Cosmologie, Univ. Grenoble Alpes, CNRS/IN2P3\\ 53 avenue des Martyrs, 38026 Grenoble Cedex, France
}

\date{\today}

\begin{abstract}
The idea that the evaporation of black holes could end around the Planck mass and lead to relics possibly accounting for dark matter is quite old. In this article we focus on new models where the relics are metastable, typically with a lifetime of order $M^5$, where $M$ is the initial mass of the black hole. We show that those relics would actually be quite luminous, radiating -- possibly as much as the stars -- in the optical or UV band. We explicitly construct a proposal for their direct detection.
\end{abstract}

\maketitle

\section{Introduction}

This article considers the possibility that dark matter is made of metastable black (or white) hole relics in a reasonable model dealing with the information paradox.\\

The nature of dark matter is one of the most important known problems of physics, especially since supersymmetric particles are no longer viable (or, at least, natural) candidates \cite{Feng:2022rxt}. Numerous astrophysical searches are being carried out (see, {\it e.g.}, \cite{Censier:2011wd,Gascon:2015caa,Mayet:2016zxu,Cirelli:2012tf,Conrad:2014tla,Gaskins:2016cha,deDiosZornoza:2021rgw} for reviews). Using accelerators as dark matter factories is also actively considered (see, {\it e.g.}, \cite{Kahlhoefer:2017dnp,Felcini:2018osp} for reviews). Theoretical ideas are very numerous (see, {\it e.g.}, \cite{Plehn:2017fdg,Oks:2021hef,Kisslinger:2019ysx} for reviews) and new ones appear on a nearly daily basis. The observational evidence for dark matter are very convincing (although they do not fully rule out modified gravity approaches \cite{Acedo:2017dvl,Katsuragawa:2017whg,Milgrom:2019cle}) but the direct search for candidates has failed so far. From the theoretical point of view, black holes might be the best candidates as they are known to exist (see \cite{LIGOScientific:2016aoc,EventHorizonTelescope:2019dse,EventHorizonTelescope:2019ths} for recent observations which add up to historical arguments). In a sense, they require a minimal amount of ``new physics".\\

The idea that primordial black holes produced in the early universe could form dark matter is not new (see, {\it e.g.}, \cite{Khlopov:1999rs,Khlopov:2008qy} for broad introductions and \cite{Villanueva-Domingo:2021spv,Carr:2020xqk,Carr:2021bzv} for recent reviews). Recently, it was for example suggested that primordial black holes in the holographic space-time model could account for dark matter \cite{Banks:2020dgx,Banks:2021lai}. 
Those scenarios are however very severely constrained and not much parameter space is allowed.

When it is light enough, a primordial black hole of mass $M$ is expected to evaporate by the Hawking mechanism  \cite{Hawking:1974sw}, leading to a nearly thermal law characterized by a temperature $T_H=1/(8\pi M)$ (in Planck units $\hbar = c = k_B = G = 1$, used throughout this article except otherwise stated). This process is consensual and understood from quite a lot of different perspectives (see, {\it e.g.}, \cite{Lambert:2013uaa} for a simple introduction). In a way, it might even be considered as observed in analog systems \cite{Steinhauer:2015saa}.\\

The semiclassical calculation however breaks down when the mass approaches the Planck mass. Numerous arguments were given in favor of the existence of stable relics (see, {\it e.g.}, \cite{Zeldovich:1983cr,Aharonov:1987tp,Banks:1992ba,Banks:1992is,Barrow:1992hq,Bowick:1988xh,Casher:1991en,Coleman:1991jf,Lee:1991qs,Gibbons:1987ps,Torii:1993vm,Callan:1988hs,Myers:1988ze,Whitt:1988ax,Alexeyev:2002tg,Hossenfelder:2003dy,Maziashvili:2005pp,Xiang:2006mg,Nikolic:2006nd,Banerjee:2010sd,Ali:2012hp,Rubiera-Garcia:2013gnv,Lobo:2013adx,Dirkes:2013dca,Ali:2014xqa,Chen:2014jwq,Wen:2015kwa,Mehdipour:2016vxh,Li:2017kze,EslamPanah:2018rob,Ong:2018syk,Chamseddine:2019pux,Kuntz:2019gka,Khan:2020tdk} for reviews). The reasons given in favor of an interruption of the evaporation in the Planck regime  are extremely diversified and all lead to the conclusion that black holes should form remnants. One of the most convincing points, relying only on known physics, was made by Giddings in \cite{Giddings:1992hh} (in the information paradox framework, see \cite{Mathur:2012np}).

Those relics are good candidates for dark matter. The idea that dark matter could be made of black hole remnants with masses $M_{rel}\sim M_{Pl}$ was originally suggested in Ref. \cite{MacGibbon:1987my}. The kind of fluctuations that was then assumed -- a blue power spectrum -- is however not anymore compatible with data. The scenario was revived in Refs. \cite{Saini:2017tsz} and \cite{Conley:2006jg,Nakama:2018lwy} (see also references therein) for extra-dimensional models, and in Ref. \cite{Dalianis:2019asr} for runaway-quintessence postinflationary models, among quite a lot of other proposals (see, {\it e.g.}, \cite{Barrau:2003xp,Chen:2004ft}). In Ref. \cite{Barrau:2019cuo}, black hole relics were considered as a dark matter candidate relying only on known physics and, more speculatively, in Ref. \cite{Barrau:2022eag} within the holographic space-time hypothesis.

The entire idea is especially interesting as the production of larger-than-expected amounts of primordial black holes during inflation is also being seriously studied (see \cite{Dalianis:2019asr,Lehmann:2019zgt,Byrnes:2021jka,Ozsoy:2021pws,Heydari:2021qsr,Tomberg:2021xxv,Ahmed:2021ucx,Gangopadhyay:2021kmf,Flores:2021jas,Spanos:2021hpk,Teimoori:2021thk,Rigopoulos:2021nhv,Zheng:2021vda,Solbi:2021rse,Bastero-Gil:2021fac,Solbi:2021wbo,Ng:2021hll,Gao:2020tsa,Ashoorioon:2020hln,Zhou:2020kkf,Vennin:2020kng,Ragavendra:2020sop,Ozsoy:2020kat,Conzinu:2020cke,Palma:2020ejf,Ballesteros:2020qam,Lin:2020goi,Arya:2020xeo,Ashoorioon:2019xqc,Arya:2019wck,Fu:2019ttf,Bhaumik:2019tvl,Dimopoulos:2019wew,Martin:2019nuw,Kannike:2017bxn}  for some arguments).\\

In this brief article, we consider the specific -- and somehow more realistic -- case of metastable relics. We first describe the model  used in the following. Then, we show how it can account for dark matter and the related constraints that can be derived. An original way to directly detect relics is described in detail. We finally investigate several consequences and point out some important forgotten aspects.

\section{Metastable Planck relics}

The idea that relics are fully stable might be over-simplistic. Quantum gravity models suggest that they could actually be metastable \cite{Rovelli:2024sjl}. The accurate estimate of the lifetime of relics is hard and has been intensively debated \cite{Rovelli:2014cta,Barrau:2014hda,Haggard:2014rza,DeLorenzo:2014pta,Haggard:2015iya,Christodoulou:2016vny,Rovelli:2017zoa,Rovelli:2018okm,Martin-Dussaud:2019wqc,Barrau:2021spy,Kazemian:2022ihc,Dierckx:2026dky,Bianchi:2026ytw}, together with the phenomenology of dark matter in this framework. Basically, the scenario is as follows \cite{Rovelli:2024sjl}. First, a given (low mass) black hole, formed in the early universe, evaporates by the Hawking process and its mass decreases. Once it reaches the Planck region, the process ends. Quantum-
gravitational pressure stops the gravitational collapse and
induces a bounce \cite{Rovelli:2014cta}, permitting the content of the
black hole to dissipate, analogously to the cosmological bounce predicted by loop quantum gravity: the dominant effect at high density
is a quantum pressure sufficient to counterbalance weight and reverse the space collapse \cite{lqc9}. In the meanwhile, the geometry outside
the black hole, including the horizon, bounces too, tunnelling from a black to a white hole geometry \cite{Dona:2025snr}. This can be very brief in proper time but, due to the
huge gravitational time dilation, very long for an external observer. The time it takes for the remnant to disappear is a non-trivial question. The key point is that it depends on the {\it initial} mass $M$ of the black hole. This should not come as something weird: the no-hair theorem states that two black holes
with the same mass, electric charge, and angular momentum, have the same external geometry, but not that they have the same interior. As we shall explain later, older black holes, at a given mass, have a larger internal volume (in the appropriate foliation) \cite{Christodoulou:2016tuu}, hence the past history of the relic can quite naturally influence its fate from a quantum gravitational viewpoint \cite{Rovelli:2014cta}. Black holes do have a (hidden) memory!\\

To remain as generic as possible, we do not choose here a specific (and unavoidably speculative) quantum gravity scenario but, instead, try to rely mostly on general arguments based on the information paradox. The basic arguments allowing a rough but extensive characterisation of the process are quite straightforward \cite{Martin-Dussaud:2025qtr}. First, one can notice that during the evaporation from its initial mass $M$, the black hole has emitted roughly $M^2$ photons (or quanta). To preserve unitarity, a Planck-scale relic must therefore emit at least $M^2$ quanta. As the relic has a mass of order of the Planck mass, that is $M_{rel} \sim 1$ in Planck units, the mean energy per quantum is of order $E\sim 1/M^2$. Because these remnant-emitted quanta cannot be strongly correlated
with each other -- since 
they are already entangled with the earlier Hawking radiation -- each must be emitted ``independently". Roughly, the emission of each quantum takes about
the time for the wave to move out of the relic, that is $t \sim \lambda \sim 1/E \sim M^2$  
Hence, the total emission
time, that is the lifetime of the remnant, satisfies the bound $t_{rel} \geqslant M^4$. This is to be compared with the Hawking evaporation time which is of order $t_H \sim M^3$. The entire model is based on the arguments initially given in Ref. \cite{Preskill:1992tc}, to which we refer the interested reader.\\

A lifetime for the relics of order $M^4$ was precisely the hypothesis made in Ref. \cite{Rovelli:2024sjl}. If those relics are to significantly account for dark matter, the time elapsed since their formation, that is the age of the universe, must be larger than $M^3$ (so that the Hawking evaporation is completed) and lower than $M^4$ (so that the remnant has not yet disappeared) 
This basically imposes that the initial mass lies in the interval $10^{10}\,\mathrm{g}<M<10^{15}\,\mathrm{g}$. Contrarily to what has been claimed, we believe that this is however not a viable scenario as this mass range is very severely constrained \cite{Carr:2020gox,Carr:2020xqk} by the big bang nucleosynthesis and the cosmological microwave background. The mass fraction of the universe allowed to go into black holes of such masses is too small to allow remnants to account for dark matter. This initial version of the model was therefore, in our opinion, not working. \\

A new version \cite{Martin-Dussaud:2025qtr} however reaches a different conclusion. It relies on the fact that the internal volume of a black hole of initial mass $M$ scales as $tM^2$, where $t$ is the retarded time since the black hole’s
formation \cite{Christodoulou:2016tuu}. At the end of the Hawking evaporation, the volume inside the black hole is therefore of order $M^5$ (this is an approximation as the horizon shrinks in the process, but this does not affect substantially the result). This is the volume inherited by the remnant. The remnant is a ``very long" geometrical entity \cite{Rovelli:2024sjl}. During the process, the black hole has radially shrunk, but has grown in terms of its internal length.

The metric of a white hole is however the time-reversal of the one of a black hole. As the estimate of the internal volume is only based on the metric, it works also for the white hole phase. This means that the time it takes for a Planckian
white hole with a given interior volume to disappear is the same as the time it would take a Planckian black hole to develop the same interior volume. Otherwise stated, the lifetime of the remnant should therefore be of order $M^5$ \cite{Martin-Dussaud:2025qtr}.

\section{Dark matter}

This $M^5$ lifetime changes the game. The Hubble time should now be between $M^3$ and $M^5$ and the allowed mass interval (for the black holes leading to the considered relics) is therefore widened to $10^{7}\,\mathrm{g}<M<10^{15}\,\mathrm{g}$. This makes a huge difference as the ``new" range $\left[ 10^7\,\mathrm{g}\, , 10^9\,\mathrm{g} \right]$ 
is precisely the one relevant for dark matter, in the sense that those black holes can be numerous enough to create relics accounting for the entire dark matter without being in conflict with the usual cosmological scenario. The tremendous amount of radiation injected in the Universe by the Hawking evaporation remains fully compatible with all known data. \\

The question to be addressed can be summarized as follows. The mass fraction of the Universe collapsing into primordial black holes at their formation epoch is denoted as $\beta (M)$. One often refers instead to $\beta '(M)$, which accounts for the combination $\gamma ^{1/2} g_{*i}^{-1/4}
 h^{-2}$ but is of the same order of magnitude as $\beta(M)$ ($\gamma$ is related to the formation mechanism and is of order one, $g_{*i}$ is the number of relativistic degrees of freedom $10^{-5}\,\mathrm{s}$ after the Big Bang, and $h$ is the reduced Hubble parameter). In practice, $\beta'$ is not severely constrained for masses below $10^9\,\mathrm{g}$ whereas it is drastically bounded from above for $M>10^9\,\mathrm{g}$ \cite{Carr:2020gox,Carr:2020xqk}. Mostly due to the fact that black holes are less diluted than the surrounding radiation during the expansion of the Universe, their $\Omega$ parameter today (that is their density normalized to the critical density) can be much higher than the $\beta '$ at the formation time.
 In other words, a tiny amount of primordial black holes can account for all of dark matter now. Obviously the contribution of relics is suppressed when compared with the one of stable black holes. Whereas it reads $\Omega \sim 10^{9}\beta (M/M_{\odot})^{-1/2}$ for the latter \cite{Carr:2020xqk}, where $M_{\odot}$ is the mass of the sun, it becomes
 \begin{equation}
     \Omega \sim 10^{9}\beta \left( \frac{M}{M_{\odot}} \right)^{-\frac{1}{2}}\left(\frac{M_{pl}}{M}\right)\sim 5\times 10^{21}\beta 
     \left( \frac{M}{1~\mathrm{g}} \right)^{-\frac{3}{2}}
 \end{equation}
for relics. It is easy to check that the known constraints on $\beta'$ do not prevent $\Omega$ from reaching $\sim 1$ as long as $M<10^9\,\mathrm{g}$ (and $M>1 \,\mathrm{g}$ which basically corresponds to the Hubble mass at the end of inflation).\\

At this stage, one can conclude that metastable relics of black holes formed in the early universe with initial masses between $10^7\,\mathrm{g}$ and $10^9\,\mathrm{g}$ seem to be good dark matter candidates for the considered model.

\section{Detection}

If those relics were fully stable, detection would be nearly hopeless. The mean density of dark matter in the Universe is of order $2\times 10^{-11} M_{Pl}$ per cubic kilometer, that is, on average, one remnant per volume of the Moon. This is extremely small. Although some ideas are on the table (see, {\it e.g.} \cite{Christodoulou:2023hyt}), finding observational evidence of particles with a very low number density, the weight of a grain of dust, and no other interaction than gravity remains a priori extraordinary difficult.\\

The fact that the model we consider predicts metastable remnants makes a huge difference. They are not fully dark anymore. The emission rate (not to be confused with the wave frequency of the emitted photons which is of order $1/M^2$) for a single relic is $f_s \sim M^{-3}$, as $M^2$ photons are emitted during a time $M^5$. Maybe counter-intuitively, the larger the initial mass, the smaller the emission rate. Numerically, the rate ranges from $10^9\,\mathrm{Hz}$ at $10^7\,\mathrm{g}$ to $10^3\,\mathrm{Hz}$ at $10^9\,\mathrm{g}$. On the other hand, the energy of the emitted quanta -- hereafter assumed to be photons -- is of order $E_{\gamma} \sim 1/M^2$. For the mass range relevant to dark matter, remnants radiate in the optical and ultraviolet bands (from $10^4\,\mathrm{eV}$ for black hole masses around $10^7\,\mathrm{g}$ to $1\,\mathrm{eV}$ for black hole masses around $10^9 \,\mathrm{g}$). It is somehow reassuring that the calculated rate roughly corresponds to the one obtained by truncating the Hawking spectrum in the following sense: assuming that a Planck-size black hole ``tries" to emit quanta at its natural rate -- that is with a Planck-time period -- but that most of the spectrum is forbidden (the region corresponding to energies higher than $M^{-2}$ is not allowed for the previously given reasons), one gets a frequency of order $M^{-3}$.\\

In the Solar System, the relic number density is substantially higher -- although still tiny when compared to lighter dark matter candidates -- at $n_{rel,\odot} \sim 2\times 10^{-5}$ km$^{-3}$. One can compare, averaged over a volume much smaller than the galactic scale but much bigger than the stellar scales, the number of photons per second emitted by stars, $N_{star}$, and by relics, $N_{rel}$. For initial masses close to $10^7\,\mathrm{g}$, our calculation shows that the ratio is of order unity (and decreases to $N_{rel}/N_{star}\sim 10^{-5}$ for $M\sim 10^9\,\mathrm{g}$)! In other words, the remnants roughly emit as much as the stars. This dark matter candidate is not that dark... One can also perform the same calculation at the galactic center, leading to comparable results since the density of stars and dark matter does not scale in a radically different way as function of the galactic radial coordinate. 

Finally, we have estimated the mean extragalactic contribution due to a homogeneous (on very large scales) distribution of relics assumed to account for dark matter. The resulting number of photons received on Earth, is around $10^8\,\mathrm{m^{-2}\cdot s^{-1} \cdot sr^{-1}}$ (for $M=10^9\,\mathrm{g}$). This is below the extragalactic background \cite{Dwek:2012nb}. However, considering lower masses increases the photon flux and might lead to a signal that should have already been detected. For lifetimes even greater than $M^5$, the flux of photons becomes smaller and can be compatible with data on a larger mass interval (\textit{e.g.} $10^7$ to $10^9\,\mathrm{g}$ for $t_{rel} \sim M^6$). It should be pointed out that even if the signal is monochromatic, which is implicit when we write $E=1/M^2$, as suggested in the model of Ref. \cite{Martin-Dussaud:2025qtr}, the spectral width at detection is anyway non-vanishing, fixed by the redshift involved in the line-of-sight integration.\\ 

It is not that easy to reach a firm conclusion at this stage. The model, as developed in Refs.  \cite{Martin-Dussaud:2025qtr,Kazemian:2022ihc}, assumes that all emitted quanta have the same energy. This is, by far, its weakest hypothesis. 
If the emission is indeed monochromatic, or nearly so, it would obviously have been seen in the galaxy. If the spectral width is however quite large and the initial masses of black holes close to the higher bound, it would most probably have been unnoticed. In any case, the fact is that the light radiated by the metastable remnants is {\it not} negligible.\\

Let us now set those constraints aside and ask ourselves if metastable relics could in principle be directly detected. As a (slightly) optimistic benchmark compatible with the previous estimates, let us consider remnants emitting optical photons at a rate of $10^8\,\mathrm{Hz}$. We basically deal with tiny and rare (one per $10^5\,\mathrm{km}^3$) objects radiating, each, a power of $10^{-10}$ W, hundreds of thousands times less than a firefly. The task of seeing them is hard but maybe not impossible. The most promising way to address this question is probably to consider neutrino detectors \cite{Li:2025tqf} like KM3NeT (see Ref. \cite{KM3NeT:2024paj} and references therein) or IceCube (see Ref. \cite{IceCube:2024pnx} and references therein). A huge cubic kilometer network of photomultipliers operating in the dark might indeed be suitable for the purpose. 

It is possible to make a rough estimate of the viability of this approach. Let us first estimate the order of magnitude of the distance $R$ from a photomultiplier at which a remnant can be ``seen". The number of photons received by the photomultiplier during a time $t$ is 
\begin{equation}\label{eq: S}
    S=f_s \cdot t \cdot \frac{\pi r^2}{4\pi R^2} \cdot Q,
\end{equation}
where $r<R$ is the photocathode radius and $Q\simeq 0.25$ is the quantum efficiency \cite{Aiello_2025}. The time spent by a relic whose speed relative to the Earth is $v$ within the ``sphere of detectability" is of order $t\sim R/v$. The number of photons therefore reads $S\sim f_sRr^2Q/(vR^2)$. The number of photons of noise, on the other hand, is on average
\begin{equation}
    B=f_bt\sim f_b \cdot \frac{R}{v},
\end{equation}
where $f_b$ is the frequency of the noise, including both the background light and the dark current of the photomultiplier ($f_b\simeq 10\,\mathrm{kHz}$ in KM3NeT \cite{Aiello_2025}). The noise is mostly well-known and detection is therefore limited by statistical fluctuations of the noise. We then require 
\begin{equation}
    \frac{S}{\sqrt{B}}>x,
\end{equation}
where $x$ is somewhere between 1 and 5 depending on the expected significance. This condition translates into:
\begin{equation}
\label{eq:Rmax}
    R<\left( \frac{Qf_sr^2}{x\sqrt{vf_b}} \right)^{\frac{2}{3}},
\end{equation} 
which fixes the distance from a photomultiplier at which a remnant can be seen.\\

Furthermore, it is possible to give an estimate of the expected signal and background rates. By ``signal" we mean here a relic generating at least $K$ pulses per photomultiplier in several aligned photomultipliers (when passing through their respective spheres of influence), with the appropriate time pattern. By ``background", we mean the same kind of event generated randomly from the noise of the photomultipliers. We ignore the subtleties of the detector geometry (as there are, in practice, several photomultipliers in each detecting station). 
Using Eq. \eqref{eq: S} with $S=K$ and taking $t \sim R/v$ allows to express the radius of the spheres within which the relic leads to a signal event, at fixed relic velocity:
\begin{equation}
\label{eq:newR}
    R = \frac{r^2Q f_s}{4 K v}.
\end{equation}

Let us now consider a cube of size $L$, representing the full volume of the detector filled by $N$ spheres of effective radius $R < L$ and cross section $\sigma = \pi R^2$. Numerically, we take $N\simeq 2500$, $L\simeq 200\,\mathrm{m}$, $r\simeq 44\,\mathrm{cm}$ \cite{Aiello_2022}.
The mean length of a ``random" straight line intersecting the cube is of order $ L$. The mean number of intersections between a straight line going across a cubic detector of size $L$ and spheres of effective radius $R$ (as calculated previously) is of order
\begin{equation}
    \mu_s = n \sigma l = \frac{N}{L^3} \times \pi R^2 \times L = \frac{N \pi R^2}{L^2}~.
\end{equation}
The number of intersections $X$ between the random straight line, representing the path of the relic, and a sphere of effective radius $R$, representing the zone of possible detection by a single photomultiplier, follows a Poisson law of parameter $\mu_s$:
\begin{equation}
    P(X=k) = \frac{\mu_s^k}{k!} e^{-\mu_s}.
\end{equation}

This allows to straightforwardly calculate the probability of interacting with an arbitrary number of photomultipliers. It is reasonable to ask that 3 photomultipliers are hit so that the expected linearity of the trajectory is ensured. Should the reconstructed trajectory be directed toward the Cygnus constellation -- which corresponds to the motion of the Earth with respect to the dark matter halo -- this would be additional evidence that the luminous moving object is indeed a dark matter ``particle" or relic.

The signal rate for $k=3$ is then given by
\begin{equation}
\label{eq:signalrate}
    r_s = n_{rel,\odot}  v \cdot L^2 \cdot P(3) \sim \frac{\pi^3}{3!}    \frac{N^3  R^6}{L^4} n_{rel,\odot}  v\ ,
\end{equation}
where $\mu_s \ll 1$ is used.

This signal rate $r_s$ can be compared to the rate $r_b$ at which a false detection -- noise mimicking a signal -- is expected to happen. The main idea behind this estimation is to calculate the probability, that, during a time window of order $R/v$, one records $K$ pulses in one photomultiplier, followed by $K$ pulses in two other photomultipliers, aligned with the first one. A detailed calculation is given in Appendix \ref{ap:noise estimation}.\\

For a relic emitting photons at a rate of $10^8\,\mathrm{Hz}$, with a mean velocity of $v=220\,\mathrm{km\cdot s^{-1}}$ and a minimal one\footnote{We fix this lower velocity $v_{min}$ such that only $1\%$ of the relics have a velocity inferior to $v_{min}$ (see Appendix \ref{ap:noise estimation}).} of around $80 \,\mathrm{km\cdot s^{-1}}$, we estimate that, for a threshold $K=5$, one event per day could be expected, with a signal to noise ratio $r_s/r_b$ of a hundred. Although we leave more detailed calculations for a dedicated work, this is encouraging. The fact that the statistical distribution of speeds and directions would be accessible obviously constitutes an additional point in favor of a plausible detection.

\section{Varying the lifetime}

As mentioned before, there are good arguments to assume that the lifetime of the metastable relics is of order $M^5$. This is however not a definitive result. Some works \cite{Bianchi:2026ytw} suggest an exponential lifetime which, from the viewpoint of phenomenology, basically means that the relics behave as if they were fully stable. It is however possible that their actual lifetime scales as $M^{\alpha}$ with $\alpha$ a parameter greater than 5 (the case $\alpha <5$ has already been studied and is mostly excluded).\\

The minimum mass (obtained by requiring the lifetime of relics to be lager than the Hubble time) compatible with the dark matter hypothesis reads, as a function of $\alpha$: 
\begin{equation}
    M_{min}\sim 10^{-5}\,\mathrm{g} \times 10^{\frac{61}{\alpha}}
\end{equation}
whereas the maximum mass, fixed by the Hawking time, does not depend on the lifetime of the remnant. The maximum emission rate (that is at the minimum mass) reads
\begin{equation}
    f_s\sim 10^{44}\,\mathrm{Hz} \times 10^{61(\frac{2}{\alpha}-1)},
\end{equation}
with an associated mean energy
\begin{equation}
    E\sim 10^{28}\,\mathrm{eV} \times 10^{-\frac{122}{\alpha}}.
\end{equation}

\section{Discussion}

Several remarks are in order to underline points that seem to have been missed in the literature.

First, although the model considered here is consistent and convincing in the sense that it relies on quite basic and natural arguments, it is problematic from the full quantum gravitational viewpoint. It is indeed assumed -- for good reasons -- that the relic releases many low energy photons \cite{Kazemian:2022ihc}. This means that the horizon area takes a huge number of values between the Planck area and zero. One of the key results of loop quantum gravity is, however, that the spectrum of the area operator is discrete with the Planck area (or nearly so) as its smallest non-vanishing eigenvalue. There is a tension here. Some ways out are possible: first, the very notion of a horizon is complex in this case (there is probably no event horizon at all \cite{Rovelli:2024sjl} in this framework although the local properties of a horizon, either apparent \cite{Hawking:1973uf}, trapping \cite{Hayward:1993wb}, isolated \cite{Ashtekar:2000sz}, dynamical \cite{Ashtekar:2004cn}, do remain). Moreover, the non-commutativity of the mass and area operator in the fully quantum regime might blur the usual mass-area relation in case measurements are performed\footnote{We thank Pierre Martin-Dussaud for pointing this out.}. This is anyway something to clarify.\\

Second, a quite unique feature has to be underlined. In many classical and quantum approaches, the Big Bang singularity is replaced by a cosmological bounce (see, {\it e.g.}, \cite{Brandenberger:2016vhg,Date:2004fj,Barrau:2020nek}). Black holes (or black hole relics) are known to possibly survive the bounce \cite{Carr:2011hv} and could have been formed in the contracting branch of the universe -- the probability of their formation is then enhanced compared with that in an expanding space. 
The relics considered in this work might constitute one of the rare (if not the only) remains of what was actually created before the bounce and, even more importantly, to keep the signature of their formation epoch! If very low energy photons from dark matter remnants were detected, it would be evidence for a pre-Big Bounce formation. Let us recall that the model requires the Hawking time to be smaller than the Hubble time $t_H$ -- that is $M^3<t_H$ -- so that the relics are indeed formed as of today. As the mean energy per photon $E$ is of order $1/M^2$, this means that $E>t_H^{-2/3}$. If photons below this bound (of order $10^{-12}\,\mathrm{eV}$, i.e. wavelengths larger that radio waves) were to be observed from relics, this would inevitably mean that the black holes were formed longer ago than the Hubble time, that is, before the bounce.\\

 Although the existence of inflation does {\it not} refute the previous argument, it plays against its practical use as the black holes (or remnants) would then have been drastically diluted. The aforementioned idea is therefore especially appealing in the matter-bounce inflationless model \cite{Wands:1998yp,Finelli:2001sr,Allen:2004vz}. 

In this case, it even offers a very nice solution to the paradox of the low entropy initial state of the Universe. In Ref. \cite{Rovelli:2015dha}, it was argued that this highly peculiar (low entropy) configuration -- which is mandatory for evolution and life to exist and therefore to account for the world around us -- might be due to a perspectival effect: we use, to describe the Universe (and to interact with it through the relevant macroscopic variables), a very specific coarse-graining. In the model considered here, as pointed out in \cite{Rovelli:2018hba}, the remnants offer a concrete example of the perspectivist entropy in the sense that the ``initial" cosmological configuration is no longer a very unlikely homogeneous gravitational field but, instead, a highly crumpled geometry -- which is thermodynamically  much more favored. We, observers, are special by living in the small -- but entropically favored -- portion of the full volume which is out of the remnants at the bounce time. Increasing the initial estimate of an interior volume of order $M^4$ (given in \cite{Rovelli:2018hba}) to a volume of order $M^5$ (given in \cite{Martin-Dussaud:2025qtr}) makes the scenario obviously even more entropically appealing. A clear conclusion is obviously way beyond the scope of this brief note. 

\section{Conclusion}

The main results of this article are the following:
\begin{itemize}
    \item increasing the lifetime of remnants from $M^4$ to $M^5$ changes drastically the relevance of the model for dark matter. It becomes viable with this new estimate.
    \item those relics are actually not dark and might emit a considerable amount of light.
    \item the direct detection of a single relic is, in principle, possible thanks to neutrino detectors.
    \item there is a tension between the usual spectrum of the area operator of loop quantum gravity and the construction of the model.
    \item it might be possible, in this framework, to have a clear indication of the existence of pre-bounce (``older than the universe") astrophysical objects. 
    \item a link with the low-entropy state of the Universe could even be built but this point would deserve much deeper investigation. 
\end{itemize}

Obviously, there are many directions in which this study could be pushed forward. The most important point to clarify now, before any future phenomenological investigation, is probably to make more reliable the theoretical estimate of the actual lifetime, beyond heuristical arguments.

\appendix

\section{Background Estimation}\label{ap:noise estimation}

Although this article deals only with orders of magnitude, the question of the calculation of the frequency of background events is not mathematically evident. As we believe it can be useful for other studies, we detail here the reasoning.\\

We model the noise in a photomultiplier as a continuous
Poisson process with rate $f_b$: $\forall t>0$, the number of pulses in any time interval of size $t$ is given by a Poisson distribution $\mathcal{P}(\mu(t))$ of parameter $\mu(t) = f_b t$. We consider three independent Poisson processes $N_{i=1,2,3}(t)$ with the same rate $f_b$ (corresponding to the datafeed of three aligned photomultipliers), scanned during a large time period $[0,T]$. We fix a window size $\tau \ll T$, a threshold $K\geqslant 3$, and denote $P = e^{-\mu(\tau)} \mu(\tau)^K / K!$. We consider the following events:
\begin{enumerate}
    \item Fix a time $t_1> 0$. $E_1$: \{the first process generates $K$ pulses in the window $[t_1,t_1+\tau]$\}.
    \item $E_2$: \{there exists a window $[t_2,t_2+\tau]$, with $0\leqslant t_2 < T-\tau$, inside which the second process generates at least $K$ pulses.\}
\end{enumerate}

The {\it rate} $r_1$ at which event $E_1$ occurs as $t_1$ scans through a large time period can be obtained by the following argument: the window $[t_1-\dd t, t_1+\tau-\dd t$] must contain $K-1$ pulses (which has {\it probability} $e^{-\mu(\tau)} \mu(\tau)^{K-1} / (K-1)!$), and a new pulse  must appear in the window as it shifts (which, on average, happens at a {\it rate} $f_b$), hence 
\begin{equation}
    r_1 = f_b P K/\mu(\tau) = P K / \tau .
\end{equation}
In the language of queuing theory, this corresponds to a M/D/$\infty$ queue.

Determining the probability $P_2$ of $E_2$ is generally a difficult problem, so we rely on the approximation developed in Ref. \cite{naus_scan_statistics} (see references therein for other approximations). Define $\lambda = \mu(\tau)=f_b \tau, p_j = e^{-\lambda} \lambda^j / j!$ for $j\in\mathbb{N}$, 
$F_n = \sum_{j=0}^n p_j$ for $n\in\mathbb{N}$ and $F_n = 0$ for $n<0$. Then
\begin{equation}
    P_2 \simeq 1 - Q_2 \left(\frac{Q_3}{Q_2}\right)^{T/\tau -2},
\end{equation}
where
\begin{align}
    Q_2 &= F_{K-1}^2 - (K-1)p_K p_{K-2} - (K-1-\lambda)p_K F_{K-3}, \\
    Q_3 &= F_{K-1}^3 - A_1+A_2+A_3-A_4, \\
    A_1 &= 2 p_K F_{K-1}((K-1)F_{K-2} - \lambda F_{K-3}), \\
    A_2 &=  \frac{1}{2} p_K^2((K-1)(K-2)F_{K-3}\nonumber \\
    &\qquad\qquad - 2(K-2)\lambda F_{K-4} + \lambda^2 F_{K-5}), \\
    A_3 &= \sum_{j=1}^{K-1} p_{2K-j}F_{j-1}^2, \\
    A_4 &= \sum_{j=2}^{K-1} p_{2K-j} p_j ((j-1)F_{j-2} - \lambda F_{j-3}) .
\end{align}
The accuracy of this approximation is discussed in \cite{naus_scan_statistics} and is excellent to our purposes.

To estimate the rate of noise events that may be falsely interpreted as signal, we perform the following analysis on three aligned photomultipliers $D_{1,2,3}$ :
\begin{enumerate}
    \item We assume the velocity $v$ of the relic to be unknown but reasonably constrained within some interval $[v_{min},v_{max}]$ (e.g. with a flat or Gaussian prior). This gives us a typical time spent by a relic in the vicinity of one multiplier, $\tau \sim R/\langle v \rangle$ (using Eq.~\eqref{eq:newR} for $R$), and an upper bound on the time a relic can spend in the detector, $T \sim L/v_{min} \gg \tau$.
    \item We scan the data of $D_1$ until we find some instant $t_1$ such that the window $[t_1,t_1+\tau]$ contains $K$ pulses.
    \item We then look at the data of $D_2$ and find whether, in the interval $I = [t_1+ d_{12}/v_{max}, t_1+\tau+d_{12}/v_{min}]$ (modeling our ignorance of the relic velocity), there exists a second window of the same size $\tau$ containing at least $K$ pulses. Here, $d_{12}\sim L$ is the distance between $D_1$ and $D_2$, so $I\simeq [t_1,t_1+T]$.
    \item If such second window is found at time $t_2$, we look at the data of $D_3$ to find whether there are at least $K$ clicks in the third window $[t_2+(t_2-t_1)d_{23}/d_{12}, t_2+(t_2-t_1)d_{23}/d_{12}+\tau] \simeq  [2 t_2 -t_1, 2t_2-t_1+\tau]$, where $d_{12}/(t_2 - t_1) \sim v$ has given us an estimate of the relic velocity $v$.
\end{enumerate}

Given the previous computations, the rate at which this analysis leads to a ``false positive" is therefore
\begin{equation}
    r_3 = r_1 P_2 P' \simeq r_1 P_2 P = P^2 P_2 K/\tau,
\end{equation}
where we approximate the probability $P'$ of finding {\it at least} $K$ clicks for $D_3$ to the probability $P$ of finding $K$ pulses, given that for realistic values $\mu(\tau) \ll 1$. 

The final step is to multiply $r_3$ by the number of aligned triplets of photomultipliers in the detector. As discussed in the main text, we model them by $N$ spheres of radius $R$ uniformly distributed in a box of size $L\gg R$. With $N(N-1)$ choices to select $D_1$ and $D_3$, $D_2$ is aligned up to precision $R$ with the line joining $D_1$ and $D_3$ if it lies within a cylinder of radius $2R$ around that line. With a typical distance $(2/3) L$ between $D_1$ and $D_3$ \cite{robbins_constant}, there are therefore $N_{\text{3PM}} \sim N(N-1)(N-2) 8\pi/3 R^2 L / L^3 \sim (8\pi/3) N^3 R^2 / L^2$. Hence
\begin{equation}
    r_{b} \sim \mathcal{O}(1) \frac{N^3 R v P^2 P_2 K}{L^2}.
\end{equation}
This can be compared to the signal event rate  \eqref{eq:signalrate} computed in the main text. 

Numerically, we take $N\simeq 2500$, $L\simeq 200\,\mathrm{m}$, $r\simeq 44\,\mathrm{cm}$ \cite{Aiello_2022}, $\langle v \rangle = 220\, \mathrm{km\cdot s^{-1}}$  and $v_{min} = 80\, \mathrm{km\cdot s^{-1}}$. We consider the dark matter particle velocities $\vec{v}$ in the Solar System frame to be distributed according to a three-dimensional Gaussian with mean $v_{\odot} \simeq 220\, \mathrm{km\cdot s^{-1}}$ and dispersion $\sqrt{\langle \vec{v}\,^2\rangle} = v_{\odot} / \sqrt{2}$. The value of $v_{min}$ is then obtained by the requirement $\mathbb{P}(\abs{\vec{v}} < v_{min}) = 1\%$. For the value of $r_s/r_b$ presented in the main text, we have verified that $P_2 > P$ and $T \gg \tau$, as expected.

\bibliography{refs}

\end{document}